# Exploring the Effect of Image Enhancement Techniques on COVID-19 Detection using Chest X-rays Images

Tawsifur Rahman[1], Amith Khandakar[2], Yazan Qiblawey[2], Anas Tahir[2], Serkan Kiranyaz[2], Saad Bin Abul Kashem[3], Mohammad Tariqul Islam[4], Somaya Al Maadeed[5], Susu M Zughaier[6], Muhammad Salman Khan[7], Muhammad E. H. Chowdhury*[2]

[1]Department of Biomedical Physics & Technology, University of Dhaka Dhaka-1000, Bangladesh; tawsifurrahman@bmpt.du.ac.bd,

[2]Department of Electrical Engineering, Qatar University, Doha-2713, Qatar; mchowdhury@qu.edu.qa (MEHC), amitk@qu.edu.qa (AK), yazan.qiblawey@qu.edu.qa (YQ), a.tahir@qu.edu.qa (AT), mkiranyaz@qu.edu.qa (SK)

[3]Faculty of Robotics and Advanced Computing, Qatar Armed Forces-Academic Bridge Program, Qatar Foundation, Doha-24404, Qatar; Skashem@qf.org.qa (SBAK)

[4]Dept. of Electrical, Electronics and Systems Engineering, Universiti Kebangsaan Malaysia, Bangi, Selangor 43600, Malaysia; tariqul@ukm.edu.my (MTI)

[5]Department of Computer Science and Engineering, Qatar University, Doha-2713, Qatar; s_alali@qu.edu.qa (SAM)

[6]Department of Basic Medical Sciences, College of Medicine, Biomedical and Pharmaceutical Research Unit, QU Health, Qatar University, Doha-2713, Qatar; szughaier@qu.edu.qa (SMZ)

[7]Department of Electrical Engineering (JC), University of Engineering and Technology, Peshawar, Pakistan; salmankhan@uetpeshawar.edu.pk (MSK)

## Abstract

The use of computer-aided diagnosis in the reliable and fast detection of corona virus disease (COVID-19) has become a necessity to prevent the spread of the virus during the pandemic to ease the burden on the medical infrastructure. Chest X-ray (CXR) imaging has several advantages over other imaging techniques as it is cheap, easily accessible, fast and portable. CXR images are sometimes of poor quality and so image enhancement techniques can help the machine learning models to extract valuable discriminating features from the image. Numerous works have been reported on COVID-19 detection from smaller set of original X-ray images. However, the effect of image enhancement in COVID-19 detection was not reported in the literature. This paper explores the effect of various popular image enhancement techniques and states the effect of each of them on the detection performance. We have compiled the largest X-ray dataset called COVQU-20, consisting of 18,479 normal, non-COVID lung opacity and COVID-19 CXR images. To the best of our knowledge, this is the largest public COVID positive database. Ground glass opacity is the common symptom reported in COVID-19 pneumonia patients and so a mixture of 3616 COVID-19, 6012 non-COVID lung opacity, and 8851 normal chest X-ray images were used to create this dataset. Five different image enhancement techniques: histogram equalization (HE), contrast limited adaptive histogram equalization (CLAHE), image complement, gamma correction, and Balance Contrast Enhancement Technique (BCET) were used to improve COVID-19 detection accuracy. Six different Convolutional Neural Networks (CNNs) (ResNet18, ResNet50, ResNet101, InceptionV3, DenseNet201, and ChexNet) were investigated in this study. Gamma correction technique outperforms other enhancement techniques in detecting COVID-19 from standard and segmented lung CXR images. The accuracy, precision, sensitivity, f1-score, and specificity in the detection of COVID-19 with gamma correction on CXR images were 96.29%, 96.28%, 96.29%, 96.28% and 96.27% respectively. Classification performance using segmented lungs X-ray images slightly decreased however, the reliability of network performance significantly improved, which was observed using visualization technique. The accuracy, precision, sensitivity, F1-score, and specificity were 95.11 %, 94.55 %, 94.56 %, 94.53 % and 95.59 % respectively for segmented lung images. The enhancement techniques were also compared in terms of the time taken by the network while using each of these techniques. The proposed approach with very high and comparable performance will boost the fast and robust COVID-19 detection using chest X-ray images.

---

\*     *Correspondence: mchowdhury@qu.edu.qa; Tel.: +974 3101 0775*

## 1. Introduction

Coronavirus Disease 2019 (COVID-19) pandemic with the exponential infection rate has overloaded worldwide healthcare systems [1]. Currently, there are more than sixteen million active cases and more than one million deaths in the world, as of November 2020 [2]. COVID-19 diagnosis is carried out by Reverse Transcription Polymerase Chain Reaction (RT-PCR), which suffers from low accuracy, delay and low sensitivity [1, 3, 4]. Early diagnosis of a disease increases the chances for successful treatment of infected patients and also reduces the chances of spreading in the community for a contagious disease like COVID-19. Radiography images such as chest X-ray (CXR) or Computed tomography (CT) are routine technique for diagnosing lung related diseases such as pneumonia [5], Tuberculosis[6] and can be useful in COVID-19 detection as well [7, 8]. One of the advantages of CXR is the ability to perform them easily using portable X-ray machines providing faster, accurate COVID-19 diagnosis [7, 9-11]. CXRs are found to be potential for detecting COVID-19 with the help of artificial intelligence (AI), and are also less harmful for human body compared to CT [7, 9-11].

Recently, a large number of works have been carried out to detect COVID-19 using X-ray images with the help of AI models. Ioannis et al. [12] reported 96.78 % accuracy for COVID-19 from Bacterial Pneumonia and Normal X-rays in a dataset of 1427 X-ray. Similarly, Abbas et al. [13] reported an accuracy of 95.12 % for COVID-19 classification from COVID-19, Normal and SARS CXR's using their pre-trained CNN model (DeTraC Decompose, Transfer and Compose) with a small database of 196 X-ray images. Minaee et al. [14] reported a specificity and sensitivity of 90% and 97 % respectively using ChexPert dataset [15]. These results showed the potential of using CNN to distinguish COVID-19 from other lung diseases using CXR images. Khan et al. [16] explored limited number of machine learning algorithms for a four class classification problem (COVID-19, Bacterial Pneumonia, Viral Pneumonia and Normal) with a very small dataset. Ashfar et al. [17] reported an accuracy of 95.7 % using a Capsule Networks, called COVID-CAPS rather than a conventional CNN to deal with a smaller dataset. Some researchers have created a small dataset of COVID-19 CXR images to train machine learning models for automatic COVID-19 detection [11, 18]. The datasets consist of COVID-19 X-ray and CT images taken from the published articles. Goldstein et al. [19] built a classifier to detect COVID-19 using pre-trained deep learning model (ReNet50) and enhanced by data augmentation and lung segmentation with the help of 2362 frontal CXR images collected from four hospitals in Israel and achieved 89.7 %, 87.1 % accuracy and sensitivity respectively. Wang and Wong [20] on the other hand used around 14k CXRs but reported only 83.5 % accuracy using a deep CNN, called COVID-Net. Motamed et al. [21] proposed a randomized generative adversarial network (RANDGAN) that detects images of an unknown class (COVID-19) from known and labelled classes (Normal and Viral Pneumonia) without the need for labels and training data from the unknown class of images (COVID-19) using 14,100 CXR and attaining an AUC of 0.77. Angelica et al. [22] introduced a graph based deep semi-supervised framework for classifying COVID-19 from CXR images using around 15,254 images and achieved an accuracy of 96.4 %. Degerli et al. in [23] proposed a novel method for the joint localization, severity grading, and detection of COVID-19 from 15495 CXR images by generating the so-called infection maps that can accurately localize and grade the severity of COVID-19 infection with 98.69 % accuracy. Chowdhury et al. [24] proposed an ensemble of deep CNN models named as Efficient Convolutional Network (ECOVNet) to detect and classify COVID-19, normal and pneumonia using 16,493 CXR and achieved an accuracy of 97 %. Yamac et al. [25] introduced a compact CNN architecture, Convolution Support Estimation Network (CSEN) that utilizes CheXNet as a feature extractor to classify the target CXR images as COVID-19, Bacterial pneumonia, Viral Pneumonia or Normal. The network produced 98% COVID-19 detection sensitivity using a dataset of 462 COVID-19 CXR images. Same group of researchers have proposed a reliable advance warning system to diagnose early-stage COVID-19 cases using compact CSEN network [26]. Serval deep and compact classifiers were evaluated with 155 early-stage COVID-19, and 1,579 Normal CXR images. It was reported that CheXNet and CSEN have achieved a COVID-19 detection sensitivity of 97.1% and 98.5% respectively. Ahmed et al. in [27] proposed a novel CNN architecture,

ReCoNet (residual image-based COVID-19 detection network) for COVID-19 detection using preprocessing steps, which was reported to be very useful for enhancing unique COVID-19 signature. The proposed modular architecture trained on 15,134 CXR images and achieving 97.48 %, 96.39 % and 97.53 % accuracy, sensitivity and specificity respectively. The machine learning model consists of a CNN-based multi-level preprocessing filter block in cascade with a multi-layer CNN-based feature extractor and a classification block.

There is a demand for medical image enhancement to help clinicians make accurate/proper diagnosis of disease [5, 7, 11, 28]. Image enhancement process consists of a collection of techniques that are used to improve the visual appearance of an image such as removing blur and noise of the image, which in turn increase contrast, and gives more details of an image. The photographs taken using cellular phones and smartphones are generally of poor contrast. Therefore, this type of images needs enhancement algorithms to improve its contrast. It is required to improve the overall quality of the image, which improves the spatial features of the image. The main purpose of image enhancement is to improve the interpretability or perception of information contained in the image for human viewers or feature extraction and creates an image that is subjectively better than the original image by changing the pixel's intensity of the input image. A major concern is not to alter the information during the image enhancement process. Image enhancement plays a crucial role in many fields such as medical image processing, remote sensing, high definition television (HDTV), hyperspectral image processing, microscopic imaging etc. [29]. Various image enhancement techniques such as de-noising algorithms, filtering, interpolation, wavelets etc. [30-32] are applied for this purpose. Many functions are available to enhance the geometric features such as edges, corners and ridges of the medical images. These techniques and approaches can enhance the classification performance of the machine learning models for the medical images. Several local image enhancement algorithms have been introduced in the last two decades to improve the image quality to boost machine learning models' performance [33, 34]. Arun et al. [35] proposed the adaptive histogram equalization technique, which can help in image enhancement however not free from fuzzy appearance of the image. This approach was further improved by Hasikin et al. [36], where they proposed the use of fuzzy set theory. It not only produced better quality images but also required minimum processing time. Selvi et at. [37] proposed a method for enhancing the fingerprint images. A four-step image enhancement technique, i.e. preprocessing, fuzzy based filtering, adaptive thresholding and morphological operation, was utilized for producing noise free fingerprint images. This technique produced better peak signal-to-noise (PSNR) values than many previous techniques. Mohammad et al. [38] presented Bi-and Multi-histogram methods to enhance the contrast while preserving the brightness and natural appearance of the images. This technique has been useful in many applications that require image enhancement [39]. Several other popular histogram techniques can be explored for CXR images to investigate whether they can help AI in various image classification techniques or not [7, 20, 40-47]. In our previous study [7], we have discussed four different pre-processing schemes which were tested for detecting COVID-19 from other coronavirus family diseases (Severe Acute Respiratory Syndrome (SARS) and Middle East Respiratory Syndrome (MERS)) using CXR images. It was observed that 3-channel approach (combination of original, Contrast Limited Adaptive Histogram Equalization (CLAHE) and image complementation) outperforms other enhancement techniques and achieves sensitivities of 99.5%, 93.1%, and 97% for classifying COVID-19, MERS, and SARS images. Yujin et al. [45] used Histogram Equalization (HE) and gamma correction enhancement techniques for detecting COVID-19 from CXR. They proposed a patch-based convolutional neural network approach with a relatively small number of trainable parameters for COVID-19 diagnosis and it showed 92.5% sensitivity for COVID-19. Heidari et al. [27] proposed a VGG-19 pre-trained network using histogram equalization and a new three channel approach using 8474 CXR images consisting of COVID-19, community acquired pneumonia and normal cases. The three channel approach used the two sets of filtered images from the enhanced CXR and the original images, which achieved 94.5% accuracy in classifying COVID-19 images. Most of these groups notably used a very small dataset containing only a few COVID-19 samples. This makes it difficult to generalize their findings, and cannot guarantee to reproduce the results when these models are evaluated on a larger dataset. It can be summarized that the above stated studies along with many other are relying on a limited dataset for developing and validating machine learning models. Therefore, image enhancement techniques can be investigated on a large dataset and can be compared to the original X-ray image based performance. The contributions of this paper can be explicitly stated as

1. To the best of the author's knowledge, no study has extensively studied the effect of various common chest X-ray enhancement techniques on COVID-19 CXR classification,

2. In this article, COVID-19 detection results were reported from the largest lung segmented images and compared with the plain X-ray images,

3. The outcome of this study were verified by image visualization technique to confirm the findings of the deep networks.

The remaining part of the paper is divided into the following sections: Section 2 provide the details of the various pre-trained classification models, lung segmentation models, different image enhancement and visualization techniques. Section 3 describes the methodology followed in this study, and the results of the classification performance using the original and segmented CXR images enhanced using techniques along with visualization heat map in Section 4. The paper is then concluded in Section 5.

## 2. Background

*2.1 Deep Convolutional Neural Networks*

Deep CNNs have been popularly used in image classification due to their superior performance compared to other machine learning paradigm. The networks structure automatically extract the spatial and temporal features of an image. The approach of transfer learning has been successfully incorporated in many applications. [48-50], especially where large dataset can be hard to find. Thus it opens the opportunity of utilizing smaller dataset and also reduces the time required to develop a deep learning algorithm from scratch [51, 52]. For COVID-19 detection, nine pre-trained deep learning CNNs such as ResNet18, ResNet50, ResNet101 [53], DenseNet201 [54], ChexNet [55], and InceptionV3 [56] were pre-dominantly used in the literature. ChexNet is the only network that is trained on CXR's unlike the other networks that are initially trained on ImageNet database. Residual Network (in short ResNet) with several variants, solve vanishing gradient and degradation problem [53] and learn from residuals instead of features [57]. Dense Convolutional Network (in brief DenseNet) needs a smaller number of parameters than a conventional CNN, as it does not learn redundant feature maps. The DenseNet has layers with direct access to original input image and loss function gradients. Another variant of DenseNet, ChexNet is trained and validated using a large number of CXR images [55].Inception-v3 is a CNN architecture from the inception family that makes several improvements including using label smoothing, factorized 7x7 convolutions, and the use of an auxiliary classifier to propagate label information lower down the network (along with the use of batch normalization for layers in the side head). This network scales in ways that strive to use the added computation as effectively as possible through correctly factorized convolutions and aggressive regularization [56].

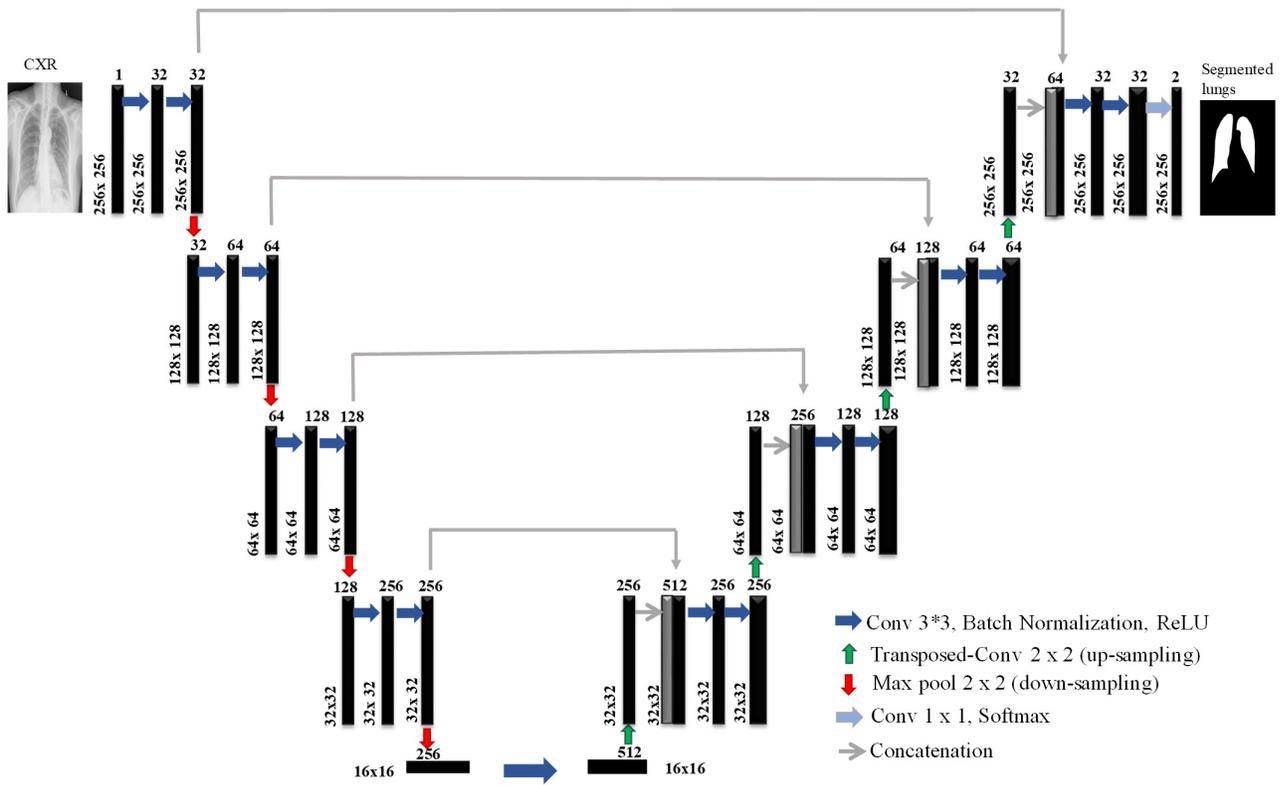

Figure 1: U-Net model architecture for lung segmentation.

*2.2 Segmentation*

Recently UNet architecture has gained increasing popularity in different biomedical image segmentation applications [58] . In this study, a U-net model with small variation in decoding part is utilized [59]. The U-net model consists of a contracting path with 4 encoding blocks, followed by an expanding path with 4 decoding blocks. Each encoding block consists of two consecutive 3x3 convolutional layers followed by a max pooling layer with a stride of 2 for down sampling. While the decoding blocks consists of a transposed convolutional layer for up sampling, a concatenation with the corresponding feature map from contracting path, and two 3x3 convolutional layer. All convolutional layers are followed by Batch normalization and ReLU activation. At the final layer 1x1 convolution is utilized to map the output from last decoding block to 2 channel feature maps, where a pixel-wise SoftMax is applied to map each pixel into a binary class of background or lung.

*2.3 Image Enhancement Techniques*

Image enhancement is an important image-processing technique, which highlights key information in an image and reduces or removes certain secondary information to improve the identification quality in the process. The aim is to make the objective images more suitable for specific application than the original images. We employ five different enhancement strategies in this research. In the following section, these image enhancement techniques will be briefly introduced:

*2.3.1 Histogram Equalization (HE)*

Histogram equalization (HE) technique aims to distribute the gray levels within an image. Each gray level is therefore equally likely to occur. HE changes the brightness and contrast of the dark and low contrast images to enhance image quality [60]. The histogram would be skewed towards the lower end of the grayscale for a dark image, and the image information would be squeezed into the dark end of the histogram. In order to create a more evenly distributed histogram, the grey levels can be re-distributed at the dark end, which can make the picture clear. The histogram of a digital image with intensity levels in the range [0, L-1] is a discrete function represented as follows:

$$h(r_k) = n_k \qquad (1)$$

Where, $r_k$ is kth intensity value, $n_k$ is the number of pixels in the image with intensity, $r_k$. Histograms are frequently normalized by the total number of pixels in the image. Assuming an M x N image, a normalized histogram is related to the probability of occurrence of $r_k$ in the image as shown in equation 2.

$$p(r_k) = \frac{n_k}{M * N} \qquad (2)$$

*2.3.2 Contrast limited adaptive histogram equalization (CLAHE)*

An improved HE variant is called Adaptive Histogram Equalization (AHE). AHE performs histogram equalization over small regions (i.e., patches) in the image, and thus, AHE enhances the contrast of each region individually. Therefore, it improves local contrast and edges adaptively in each region of the image to the local distribution of pixel intensities instead of the global information of the image. However, AHE could over amplify the noise component in the image [61]. However, images enhanced with Contrast-Limited Adaptive Histogram Equalization (CLAHE) are more natural in appearance than those produced by HE. When the HE technique was applied to the X-ray images, it was observed that it might saturate certain regions. To address this difficulty, CLAHE uses the same approach as AHE but the amount of contrast enhancement that can be produced within the selected region is limited by a threshold parameter. Firstly, the original image is converted from RGB (red, green and blue) color space to HSV (hue, saturation and value) color space as human sense color similar to HSV version. Secondly, value component of HSV is processed by CLAHE without affecting the hue and saturation. The initial histogram is cropped and each gray-level is redistributed to the cropped pixels. The value of each pixel is reduced to a user-selectable limit. Finally, the HSV processed image is re-transformed to RGB color space.

*2.3.3 Image Invert/ Complement*

The image inversion or complement is a technique where the zeros become ones and ones become zeros so black and white are reversed in a binary image. For an 8-bit gray scale image, the original pixel is subtracted from the highest intensity value, 255, the difference is considered as pixel values for the new image. For x-ray images, the dark spots turn into lighter and light spots become darker. The mathematical expression is simply:

$$y = 255 - x \qquad (3)$$

Where, x and y are the intensity values of the original and the transformed (new) images. This technique shows the lungs area (i.e., the region of interest) lighter and the bones are darker. As this is a standard procedure, which was used widely by radiologists, it may equally help deep networks for a better classification. It can be noted that the histogram for the complemented image is a flipped copy of the original image.

*2.3.4 Gamma correction*

Typically, linear operations are performed on individual pixels in image normalization, such as scalar multiplication, addition and subtraction. Gamma correction performs a non-linear operation on source image pixels Gamma correction alternates the pixel value to improve the image using the projection relationship between the value of the pixel and the value of the gamma according to the internal map. If P represents the pixel value inside the [0,255] range, $\Omega$ represents the angle value, $\Gamma$ is the symbol of the gamma value set, x is the gray scale value of the pixel (x $\in$ P). Let $x_m$ be range midpoint [0, 255]. The linear map from group P to group $\Omega$ is defined as:

$$\varphi: P \to \Omega, \Omega = \{\omega | \omega = \varphi(x)\}, \varphi(x) = \frac{\pi x}{2x_m} \quad (4)$$

The mapping from $\Omega$ to $\Gamma$ is defined as:

$$h: \Omega \to \Gamma, \Gamma = \{\gamma | \gamma = h(x)\} \quad (5)$$

$$\begin{cases} h(x) = 1 + f_1(x) & (6) \\ f_1(x) = a\cos(\varphi(x)) & (7) \end{cases}$$

where a $\in$ [0, 1] denotes a weighted factor.

Based on this map, group P can be related to Γ group pixel values. The arbitrary pixel value is calculated in relation to a given Gamma number. Let $\gamma(x) = h(x)$, and the Gamma correction function is as follows

$$g(x) = 255 \left(\frac{x}{255}\right)^{1/\gamma(x)} \tag{8}$$

where g(x) represents the output pixel correction value in gray scale.

*2.3.5 Balance Contrast Enhancement Technique (BCET)*

BCET represents a strategy for improving balance contrast by stretching or compressing the contrast of the image without altering the histogram pattern of the image data. The solution is based on the parabolic function acquired from the image data. The general parabolic functional form is defined as

$$Y = a(x - b)^2 + c \tag{9}$$

The three coefficients a, b and c are determined from the following equations using the minimum, the maximum and the mean of the input and output image values.

$$b = \frac{h^2(E - L) - s(H - L) + l^2(H - E)}{2[h(E - L) - e(H - L) + l(H - E)]} \tag{10}$$

$$a = \frac{H - L}{(h - l)(h + l - 2b)} \tag{11}$$

$$c = L - a(l - b)^2 \tag{12}$$

Where 'l' represents the minimum value of the input image, 'h' denotes the maximum value of the input image, 'e' denotes the mean value of the input image, 'L' the minimum value of the output image, 'H' denotes the maximum value of the output image and 'E' denotes the mean value of the output image.

Figure 2 shows the difference between the different image enhancement techniques.

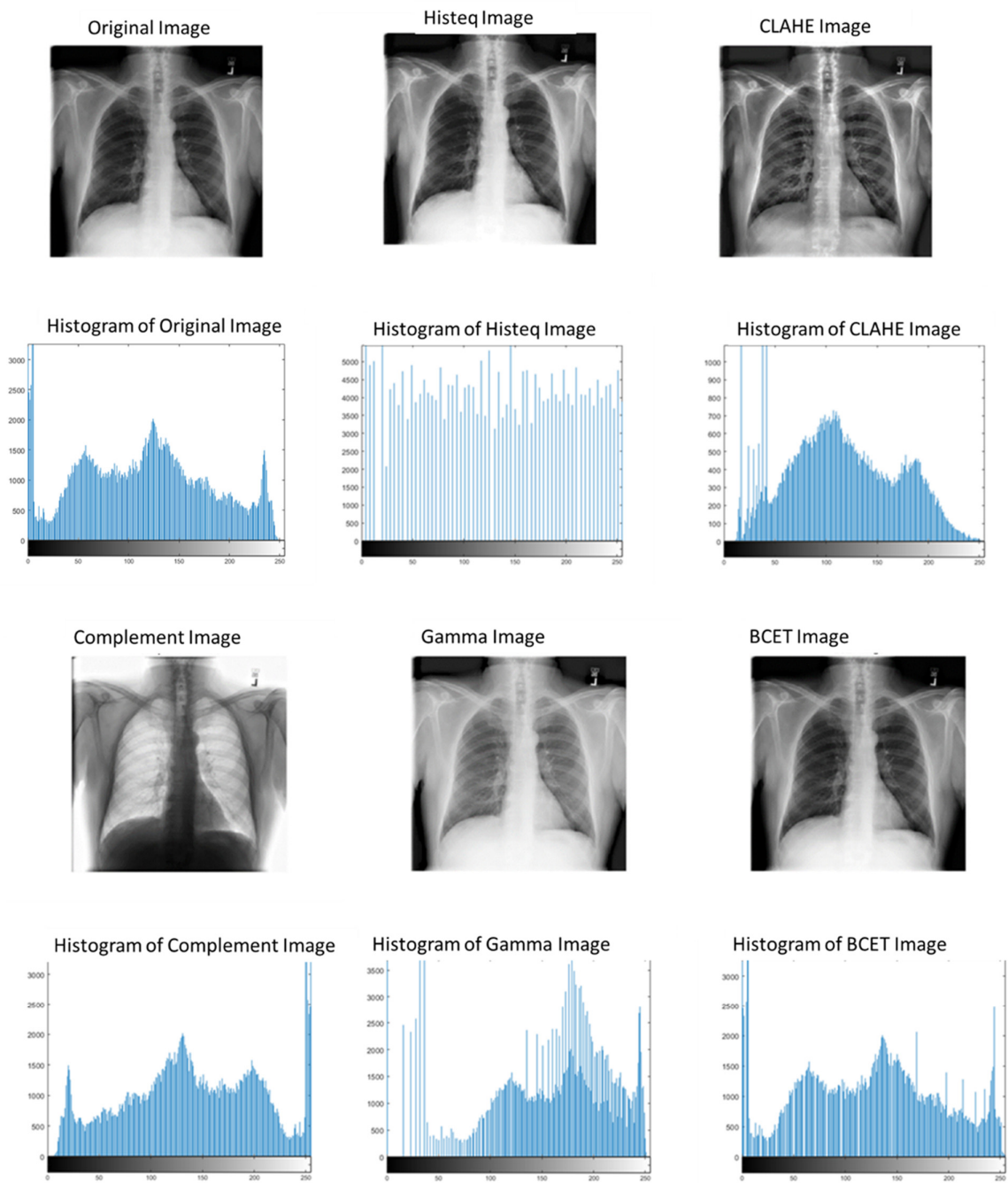

Figure 2: Histogram for original X-ray image and images undergo different enhancement techniques.

*2.4 Visualisation Techniques*

The emergence of visualization tools has led to growing interest in how CNN works and the logic behind the making of particular decisions by a network. In order to view the decision-making process of CNNs, visualization methods lead to better visual representation. These also improve the transparency of the model by visualizing the reasoning behind the inference that can be interpreted in a way that can be easily understood by humans, thereby increasing trust in the results of the CNNs. There are many popular visualization techniques such as SmoothGrad [62], Grad-CAM [63], Grad-CAM++ [64], Score-CAM [65]. But in this study Score-CAM was due to its promising performance [6]. Figure 3 provides a sample Score-CAM visualization, where it highlights the regions used by CNN in making decisions. These visualization helps in increasing the confidence of the reliability of the deep layer networks, by confirming the decision making from relevant region of the images.

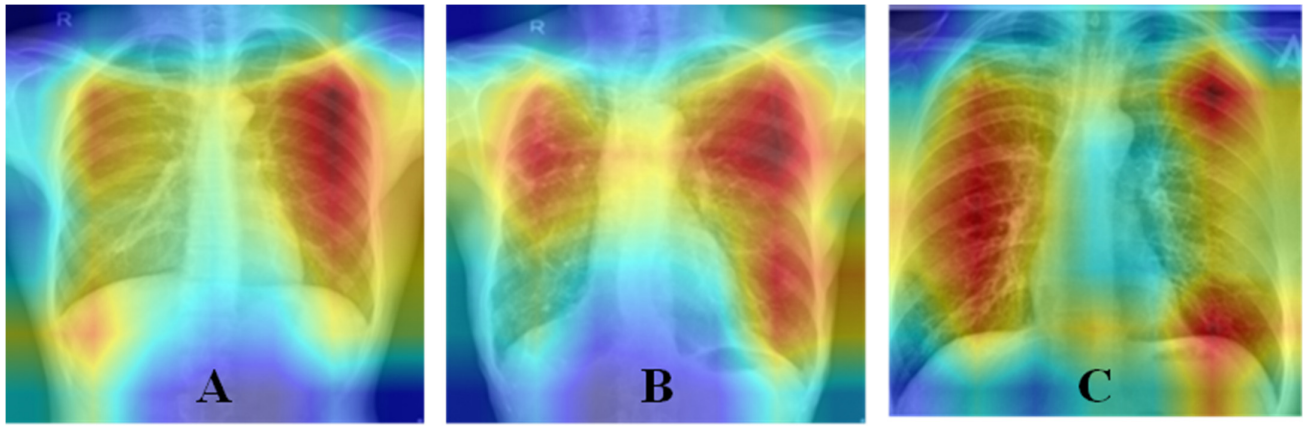

Figure 2: Score-CAM visualization of A) COVID-19 CXR, B) Normal CXR, C) Non-COVID Lung Opacity CXR, to show where the CNN model is mainly taking the decision.

**3. Methodology**

The detailed methodology adopted in the study is shown in Figure 4. The study used two different databases: 1) lung segmentation and 2) classification of CXR images into COVID-19, Non-COVID lung opacity and Normal. The major experiments that are carried out in the study: 1) Training and testing of U-Net model to segment lungs from CXR images, 2) Training and testing original and five different enhanced plain CXR images for classification using six different pre-trained networks 3) Training and testing segmented lungs of CXR images (original and enhanced) for classification using same networks. The reliability of the last two experiments were verified using Score-CAM technique.

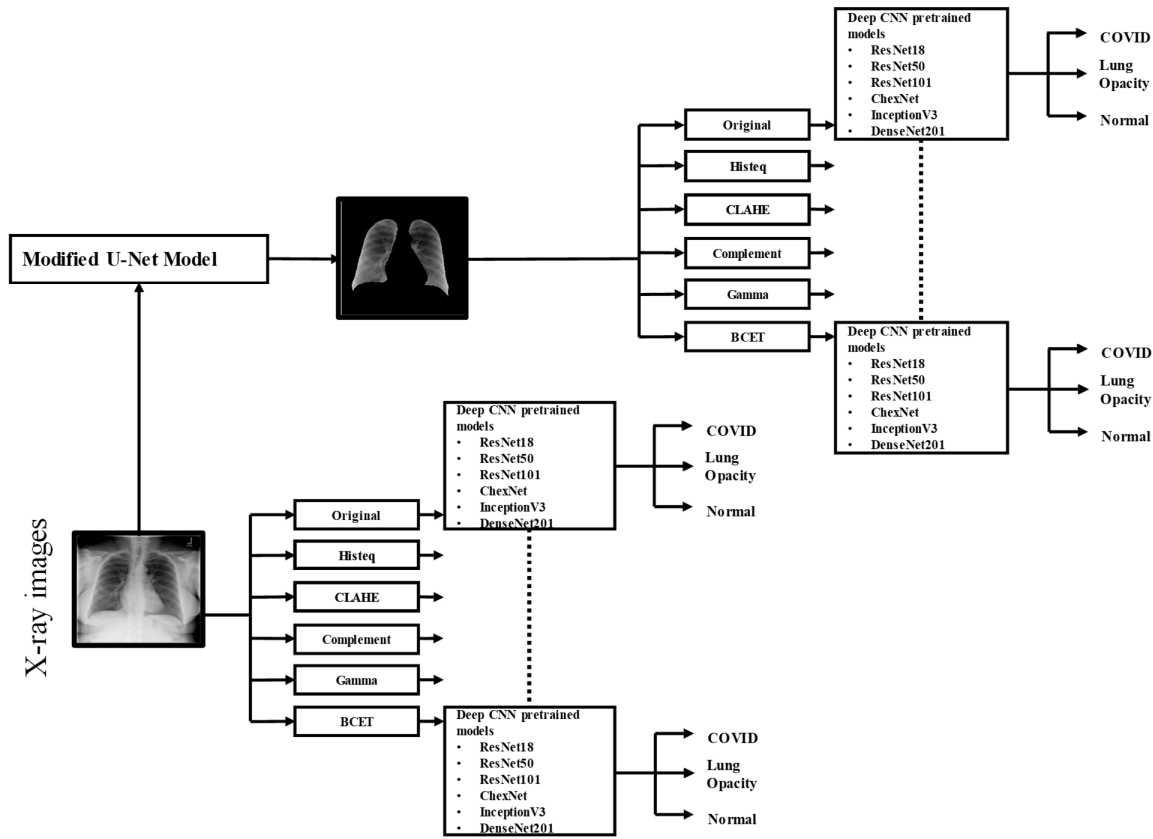

Figure 4: Block diagram of the system methodology.

The details of the study, i.e. dataset details, pre-processing and augmentation adopted in the study, performance matrices utilized in the study are discussed below.

### 3.1 Datasets Description

In this study, authors have used a large dataset compiled by the team and referred to as COVQU dataset, which is comprised of 18,479 CXR images across 15,000 patient cases

### 3.1.1 Lung Segmentation

To investigate lung segmentation models, authors have created ground truth lung masks for 18,479 CXR images which are verified by expert radiologists as a part of separate study (which is not reported in this study). Sample CXR images and masks are shown in Figure 5. The modified U-Net network was trained and tested with CXR images and their respective ground truth masks.

*3.1.2 Image Classification*

The dataset used to train and evaluate the proposed study, which is referred to as COVQU, is comprised of 18,479 CXR images across 15,000 patient cases. This COVQU dataset is the largest public COVID positive cases dataset, according to the best of the authors' knowledge. To generate this dataset, authors used and modified different open access database for three different types (COVID, Normal, and non-COVID). COVQU dataset combined RSNA CXR dataset [66] and COVID-19 dataset, details below.

The COVQU dataset compiled as a part of this study is the largest public COVID-19 positive cases dataset, according to the best of the authors' knowledge. To compile this dataset, COVID-19, Normal, and non-COVID images from different open access databases and online resources were used. Normal and Non-COVID images in COVQU dataset were taken from Radiological Society of North America (RSNA) CXR dataset [66].   In the following section, details of the RSNA and COVID-19 datasets are discussed:

*3.1.3 Image Classification*

The COVQU dataset compiled as a part of this study is the largest public COVID-19 positive cases dataset, according to the best of the authors' knowledge. To compile this dataset, COVID-19, Normal, and non-COVID images from different open access databases and online resources were used. Normal and Non-COVID images in COVQU dataset were taken from Radiological Society of North

America (RSNA) CXR dataset [66]. In the following section, details of the RSNA and COVID-19 datasets are discussed:

*3.1.4 RSNA CXR dataset*

RSNA pneumonia detection challenge dataset [66], consists of about 26,684 CXR DICOM (Digital Imaging and Communications in Medicine) images, where 8,851 images are normal, 11,821 images are with different lung abnormalities and 6,012 are non-COVID ground grass lung opacity X-ray images. In this study, we have used 8,851 normal and 6,012 non-COVID lung opacity X-ray images. Sample of the X-ray images used in the study are shown in Figure 6.

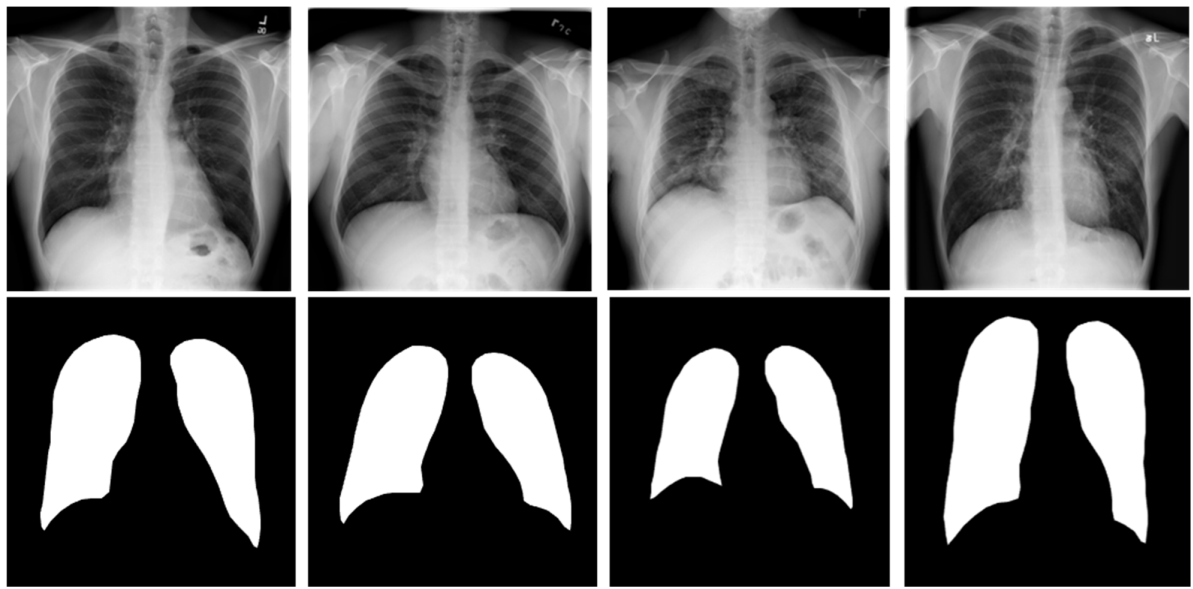

Figure 5: Samples of CXR and their ground truth masks of the dataset.

*3.1.5 COVID-19 dataset*

COVID-19 dataset is comprised of 3,616 positive COVID-19 CXR images, which are collected from different publicly available dataset, online source and published articles. Out of 3,616 X-ray images, 2,473 images are collected from BIMCV-COVID19+ dataset [67], 183 images from a Germany medical school [68], 559 X-ray image from Italian Society of Medical Radiology (SIRM), GitHub, Kaggle & Twitter [69-72], and 400 X-ray images from another COVID-19 CXR repository[73]. BIMCV-COVID19+ dataset is the single largest public dataset with 2,473 CXR images of COVID-19 patients acquired from digital X-ray (DX) and computerized X-ray (CX) machines.

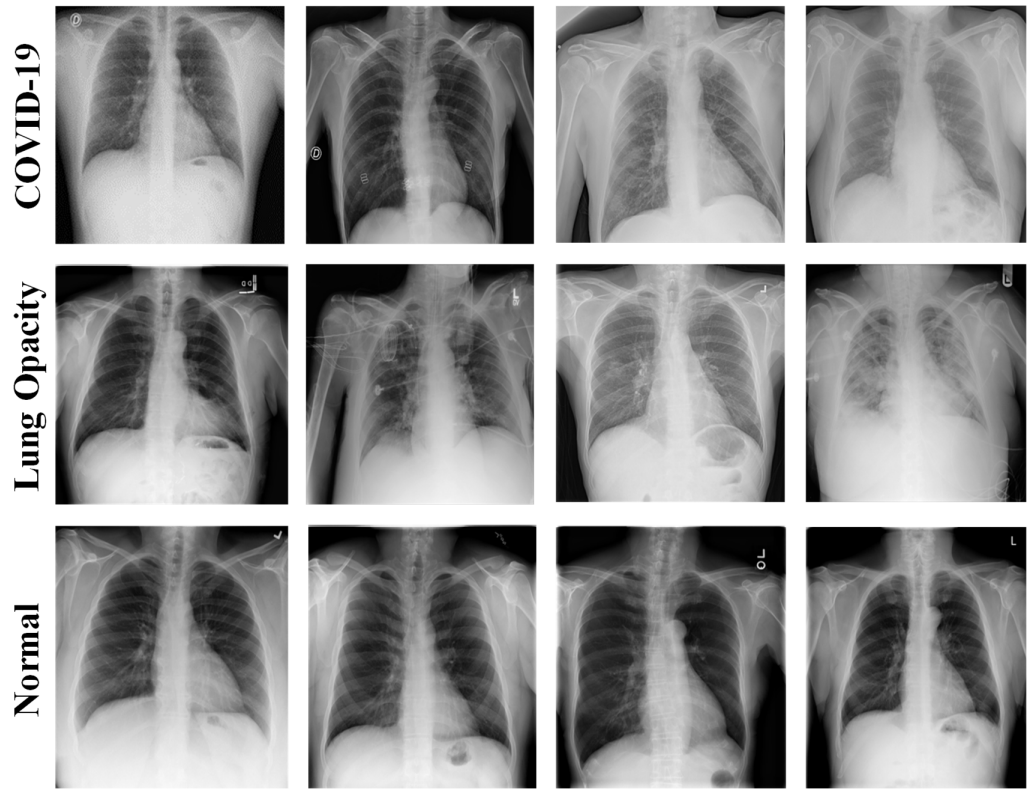

Figure 6: CXR image samples from different datasets: (A) COVID-19, (B) non-COVID Lung Opacity, (C) and Normal.

*3.2 Preprocessing And Data Augmentation*

The datasets were preprocessed to resize the X-Ray images to fit the input image-size requirements of different CNN models such as 256×256 pixels for U-Net model, 299×299 pixels for InceptionV3 and 224×224 pixels for all other networks. Using the mean and standard deviation of the images, Z-score normalization was carried out.

*3.2.1 Data Augmentation*

It is reported that the data augmentation can improve the classification accuracy of the deep learning algorithms by augmenting the existing data rather than collecting new data [74]. Data augmentation can significantly increase the diversity of data available for the training models. Image augmentation is crucial when the dataset is imbalance. In this study, the number of normal images was 8,851, which is more than twice the size of COVID-19 positive CXR images. Therefore, it was important to augment COVID-19 positive CXR images two-times to make the database balance. Some of the deep learning frameworks have built-in data augmentation facility within the algorithms, however, in this study, image rotation based augmentation technique was utilized to generate training images of COVID-19 before applying to the CNN models for training.

*3.2 Experiments*

Five-fold cross validation was used and therefore, 80 % data were used for training and 20 % for testing. Out of training dataset subset, 20 % were utilized for validation to avoiding overfitting issue [75]. Finally, the results were a weighted average of the five folds. Table 1 shows the details of the number of training, validation and test CXR images used in the two experiments of plain and segmented lung X-ray images using five different enhancement techniques.

**Table 1.** Details of the dataset used for training, validation and testing.

| Database | Types | Count. of CXR's/ class | Training Dataset | | | Test image/ fold |
|---|---|---|---|---|---|---|
| | | | Training image /fold | Augmented train image/fold | Validation image /fold | |
| COVID-19 dataset | COVID-19 | 3616 | 2314 | 4628 | 578 | 724 |
| RSNA CXR dataset | Normal | 8851 | 5664 | 5664 | 1416 | 1771 |
| | Non-COVID | 6012 | 3847 | 3847 | 962 | 1203 |

As mentioned earlier, three different experiments were conducted in this study on PyTorch library with Python 3.7 on Intel® Xeon® CPU E5-2697 v4 @ 2,30GHz and 64 GB RAM, with a 16 GB NVIDIA GeForce GTX 1080 GPU. In the following section, each of them will be discussed separately:

*3.2.1 Lung Augmentation*

Modified U-Net model was trained and validated to create lung segmentation using five-fold cross validation. Out of 18,479 CXR images and ground truth lung masks, 80% images and corresponding masks were used for training and 20% images for testing. The images were trained using batch size of 4, learning rate of 0.001, for maximum of 20 epochs using Adam optimizer. The learning rate decreased if no improvement was observed for consecutive 3 epochs and stopped if there was no improvement for consecutive 6 epochs.

*3.2.2 COVID Classification*

Six different CNN models were compared separately using non-segmented (plain) and segmented (lung) X-ray images with using five different image enhancement techniques for the classification of COVID-19, Non-COVID lung opacity, and normal images to investigate the effect of image enhancement and lung segmentation on COVID-19 detection. Five deep networks (Inceptionv3, ResNet50, ResNet101, ChexNet and DenseNet201) and one comparatively shallow networks (ResNet18) were evaluated. The images were trained using batch size of 32, learning rate of 0.001, for maximum of 20 epochs using Adam optimizer. As mentioned earlier, the learning rate was decreased with no improvement for consecutive 3 epochs and stopped with no improvement for consecutive 6 epochs.

*3.3 Performance Matrix*

*3.3.1 Lung Segmentation*

The performance of different networks in image segmentation for the testing dataset was evaluated after the completion of training and validation phase and was compared using three performance metrics: loss, accuracy, IoU, Dice. The equations used to calculate accuracy, Intersection-Over-Union (IoU) or Jaccard Index and Dice coefficient (or F1-score) are shown in equation (13-15).

$$Accuracy\ (A) = \frac{(TP + TN)}{(TP + FN) + (FP + TN)} \qquad (13)$$

$$IoU(Jaccard\ index) = \frac{(TP)}{(TP + FN + FP)} \qquad (14)$$

$$Dice\ Coefficient\ (F1 - score) = \frac{(2 * TP)}{(2 * TP + FN + FP)} \qquad (15)$$

*3.3.2 COVID Classification*

The different CNNs' performance in classification was evaluated using six performance metrics: Overall accuracy, weighted sensitivity or recall, weighted specificity, weighted precision, weighted F1 score using equations (16-20). Since different classes had different number of images, per class weighted performance metric and overall accuracy was used to compare the networks. The performance was also evaluated using area under curve (AUC):

$$Accuracy\ (A) = \frac{(TP + TN)}{(TP + FN) + (FP + TN)} \qquad (16)$$

$$Sensitivity\ (R) = \frac{(TP)}{(TP + FN)} \qquad (17)$$

$$Specificity\ (S) = \frac{(TN)}{(FP + TN)} \qquad (18)$$

$$Precision\ (P) = \frac{(TP)}{(TP + FP)} \qquad (19)$$

$$F1\ Score\ (F) = \frac{(2 * TP)}{(2 * TP + FN + FP)} \qquad (20)$$

Here, true positive (TP), true negative (TN), false positive (FP) and false negative (FN) were used to denote the number of COVID-19 CXR images were identified as COVID-19, the number of normal and non-COVID lung opacity CXRs were identified as normal and non-COVID CXRs, the number of normal and non-COVID CXRs incorrectly identified as COVID-19 CXRs and the number of COVID-19 CXRs incorrectly identified as normal and non-COVID, respectively.

In addition to the metrics stated above, the various classification networks were also compared in terms of the elapsed time per image, i.e. time taken by each network to classify an input image, represented in equation 21.

$$\delta t = t_2 - t_1 \qquad (21)$$

Where $t_1$ is the starting time for a network to classify an image, I and $t_2$ is the end time when the network has classified the same image, I.

4. **Results And Discussions**

This section describes the performance of the lung segmentation model and classification networks' performance on plain X-ray images and segmented lung X-ray images.

*4.1 Lung Segmentation*

Table 2 shows the performance matrix of the trained, validated and tested modified U-Net model for lung segmentation.

Table 2. Performance of segmentation network

| Network | Test loss | Accuracy (A) | IoU | Dice |
|---|---|---|---|---|
| Modified U-Net | 0.0376 | 98.63 | 94.3 | 96.94 |

The modified U-Net model was used to segment the classification database (8851 normal, 6012 lung opacity and 3416 COVID images), which was used for classification of COVID, lung opacity and normal cases. It is important to see on a completely unseen image-set with lung infection and normal images how well the trained segmentation model works. It can be seen from Figure 7 that the modified U-net model trained on the segmented CXR dataset can segment the lung areas of the X-ray images of the classification database very reliably. However, there is quantitative evaluation on the classification dataset is not possible as there is no ground truth masks available for this database and therefore, qualitative evaluation were done to confirm that each X-ray image was segmented correctly.

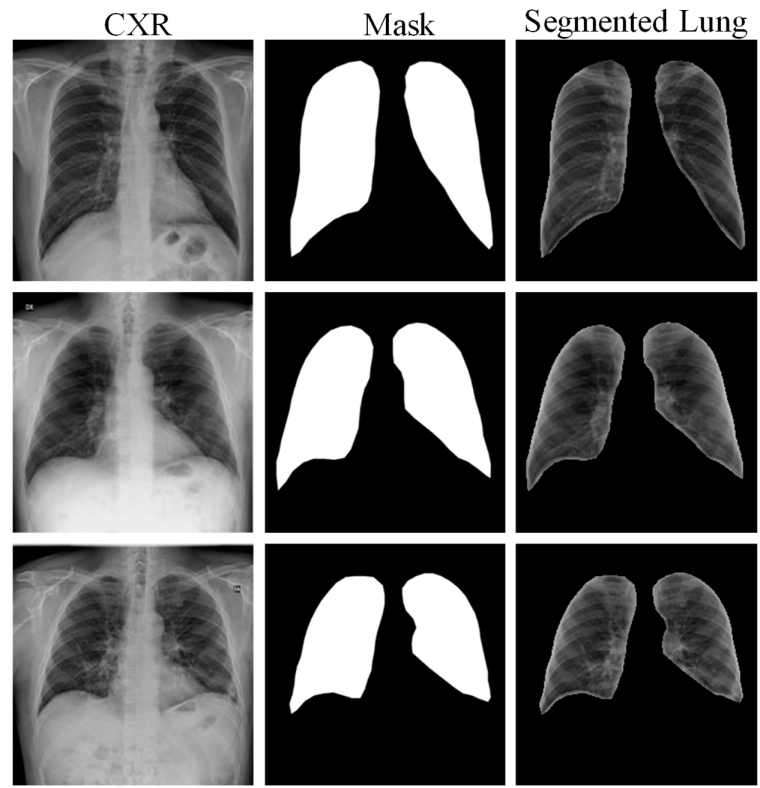

Figure 7. CXR sample images (left), generated masks by the network (middle) and resulting segmented lung (right).

*4.2 COVID-19 Classification*

As mentioned earlier, there are two different experiments (on plain and segmented lungs X-ray images) were conducted for classification of COVID-19, Non-COVID Lung opacity and Normal. The comparative performance of the best performing model for different enhancement techniques for classification between the three classes of plain images is shown in Table 3, & Table 4 shows the comparative performance of the different CNN models in classifying COVID-19 using original and Gamma corrected X-ray images.

Table 3. Comparison of the best network for the COVID-19 classification using CXR images for different enhancement techniques

| Different Enhancement | Model | Overall | Weighted | | | | |
|---|---|---|---|---|---|---|---|
| | | A | P | R | F | S | $\delta t$ |
| Original | InceptionV3 | 93.46 | 93.49 | 93.47 | 93.47 | 95.48 | 0.98 |
| Complement | DenseNet201 | 94.19 | 94.21 | 94.19 | 94.19 | 95.78 | 0.72 |
| Histeq | ChexNet | 94.34 | 94.17 | 94.14 | 94.14 | 95.98 | 0.62 |
| CLAHE | DenseNet201 | 94.08 | 94.09 | 94.08 | 94.07 | 95.77 | 0.75 |
| **Gamma** | **ChexNet** | **96.29** | **96.28** | **96.29** | **96.28** | **97.27** | **0.6** |
| BCET | DenseNet201 | 94.5 | 94.5 | 94.5 | 94.49 | 96.25 | 0.8 |

Table 4. Comparison of different models for Covid-19 classification using Original and Gamma corrected CXR images

| Technique | Model | Overall | Weighted | | | |
|---|---|---|---|---|---|---|
| | | A | P | R | F | S |
| Original | Resnet18 | 93.43 | 93.43 | 93.43 | 93.42 | 95.49 |
| | Resnet50 | 93.01 | 93.12 | 93.02 | 93.04 | 95.5 |
| | Resnet101 | 93.01 | 93.04 | 93.01 | 93 | 95.11 |
| | ChexNet | 93.21 | 93.28 | 93.21 | 93.2 | 95.54 |
| | DenseNet201 | 92.7 | 92.78 | 92.7 | 92.72 | 95.35 |
| | **InceptionV3** | **93.46** | **93.49** | **93.47** | **93.47** | **95.48** |
| Gamma | Resnet18 | 94.63 | 94.64 | 94.62 | 94.6 | 95.92 |
| | Resnet50 | 94.56 | 94.58 | 94.56 | 94.53 | 95.81 |
| | Resnet101 | 94.93 | 94.94 | 94.93 | 94.92 | 96.2 |
| | **ChexNet** | **96.29** | **96.28** | **96.29** | **96.28** | **97.27** |
| | DenseNet201 | 95.05 | 95.06 | 95.05 | 95.05 | 96.55 |
| | InceptionV3 | 94.95 | 94.95 | 94.95 | 94.93 | 96.24 |

In Table 3, the best performing network has been reported for the different enhancement techniques. It is evident that Gamma enhancement technique was the best performing enhancement technique not only in terms of classification but also in terms of elapsed time as shown in Table 3. It was further verified in Table 4, where this enhancement technique had consistently improved the performance for different network using the original X-ray images. Finally, it was seen that the combination of gamma enhancement and ChexNet was the best performing networking for the COVID-19 classification with about 96.29 % and 96.28 %, accuracy and F1-Score respectively. The superior performance of CheXNet in comparison to DenseNet201 is exhibiting that deeper layer does not always perform better and it is important to adjust hyper parameters for a specific application. It can also be noted that CheXNet is the only DenseNet variant which was initially trained on chest X-ray images. Similar performance was observed by the authors in their other COVID-19 classification problem [11]. However, ResNet18, 50 and 101 showed increasingly better performance for the classification of images without segmentation. Figure 8(A) clearly shows that the ROC curves for the best performing networks for the different enhancement techniques and it is evident that the gamma enhancement is helping the network in discriminating the different classes better. The comparative performance of the different image enhancement techniques for the different CNNs were shown in Table 5. Table 6 shows the comparative performance of the different CNN models for the classification of the three-class problem using original and Gamma corrected lung segmented X-ray images.

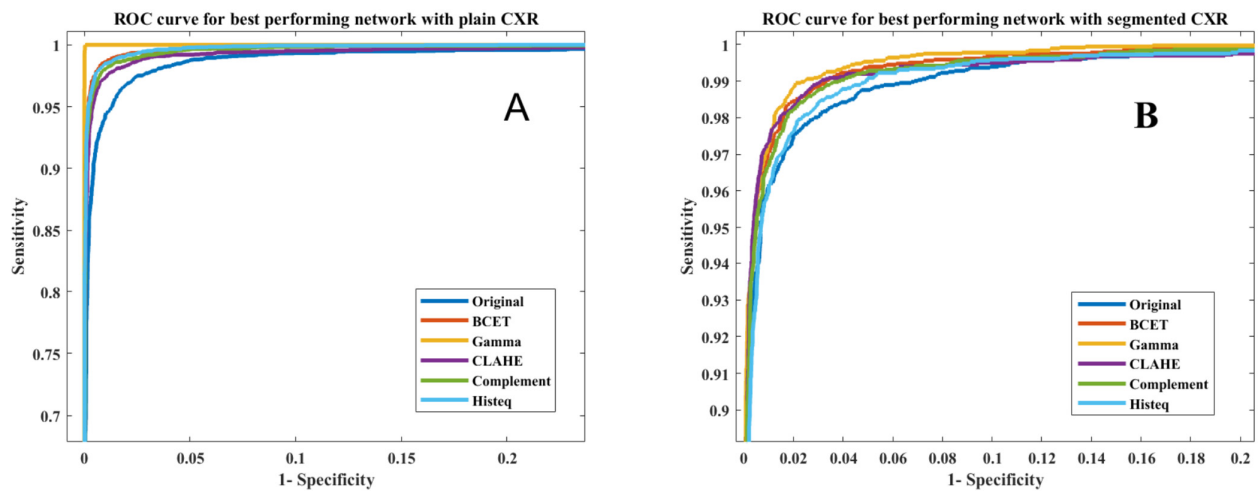

Figure 8. ROC curves for the best performing network in each image enhancement technique for plain CXR images (A) and for segmented lung CXR images (B).

Table 5. Comparison of the best network for the COVID-19 classification using segmented CXR images for different enhancement techniques

| Enhancement techniques | Model | Overall | Weighted | | | | |
|---|---|---|---|---|---|---|---|
| | | A | P | R | F | S | $\delta t$ |
| Original | ChexNet | 93.22 | 93.22 | 93.22 | 93.22 | 95.51 | 0.65 |
| Complement | InceptionV3 | 93.46 | 93.49 | 93.47 | 93.47 | 95.48 | 1.2 |
| Histeq | DenseNet201 | 93.44 | 93.43 | 93.44 | 93.42 | 95.55 | 0.78 |
| CLAHE | ChexNet | 93.9 | 93.91 | 93.9 | 93.89 | 95.77 | 0.7 |
| **Gamma** | **DenseNet201** | **95.11** | **94.55** | **94.56** | **94.53** | **95.59** | **0.72** |
| BCET | DenseNet201 | 94.12 | 94.17 | 94.14 | 94.14 | 95.98 | 0.85 |

Table 6. . Comparison of different models for Covid-19 classification
using Original and Gamma Corrected CXR images

| Technique | Model | Overall A | Weighted | | | |
|---|---|---|---|---|---|---|
| | | | P | R | F | S |
| Original | Resnet18 | 92.23 | 92.23 | 92.22 | 92.21 | 94.66 |
| | Resnet50 | 92.51 | 92.5 | 92.51 | 92.5 | 95.38 |
| | Resnet101 | 93.14 | 93.14 | 93.15 | 93.12 | 95.41 |
| | **ChexNet** | **93.22** | **93.22** | **93.22** | **93.22** | **95.51** |
| | DenseNet201 | 92.79 | 92.87 | 92.79 | 92.77 | 94.62 |
| | InceptionV3 | 92.43 | 92.44 | 92.43 | 92.4 | 94.81 |
| Gamma | Resnet18 | 93.31 | 93.3 | 93.31 | 93.3 | 95.82 |
| | Resnet50 | 93.24 | 93.24 | 93.24 | 93.22 | 95.23 |
| | Resnet101 | 93.13 | 92.92 | 92.94 | 92.92 | 95.25 |
| | ChexNet | 93.63 | 93.67 | 93.63 | 93.59 | 95.34 |
| | **DenseNet201** | **95.11** | **94.55** | **94.56** | **94.53** | **95.59** |
| | InceptionV3 | 93.92 | 93.92 | 93.92 | 93.9 | 95.7 |

Table 5 shows that Gamma enhancement technique was the best performing enhancement technique using the segmented lung X-ray images. Figure 8(B) shows the ROC curves for the different image enhancement techniques for best performing network where the enhancement technique had consistently improved the performance for different networks in comparison to the original X-ray images. It can also be seen that Gamma enhancement technique is the best performing with comparable elapsed time per image ($\delta t$) as shown in Figure 9. Finally, it was seen that the combination of gamma enhancement and DenseNet201 was the best performing networking for the COVID-19 classification with about 95.11 % and 94.56 %, accuracy and F1-Score respectively.

*4.3 Visualization Using Score-Cam*

As mentioned earlier, it is important to see where network is learning from the CXR images. It can be learning from relevant and non-relevant areas of the CXR images for classification which were verified using Score-CAM based heat maps generated for original (non-segmented) and segmented CXR images.

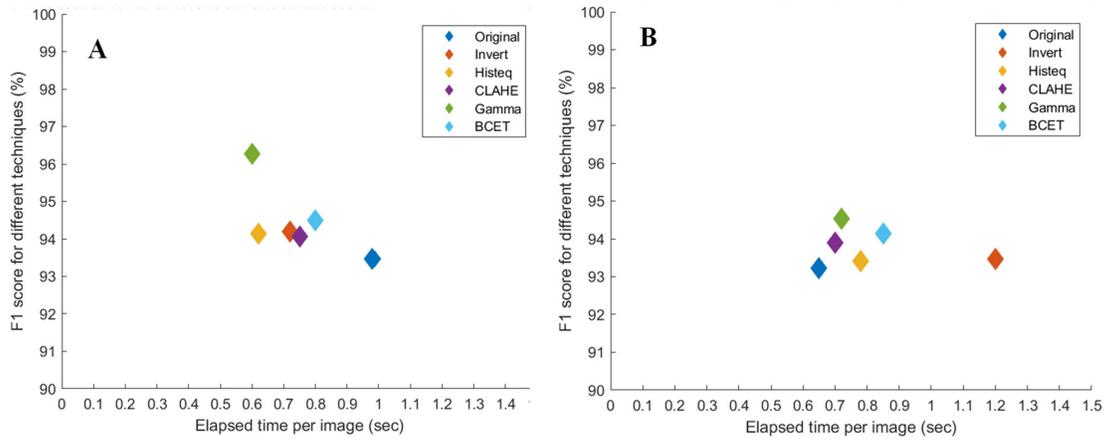

Figure 9. Comparison of F1 Score versus the elapsed time per image for the best performing network in each image enhancement technique for plain X-ray images (A) and segmented lung X-ray images (B).

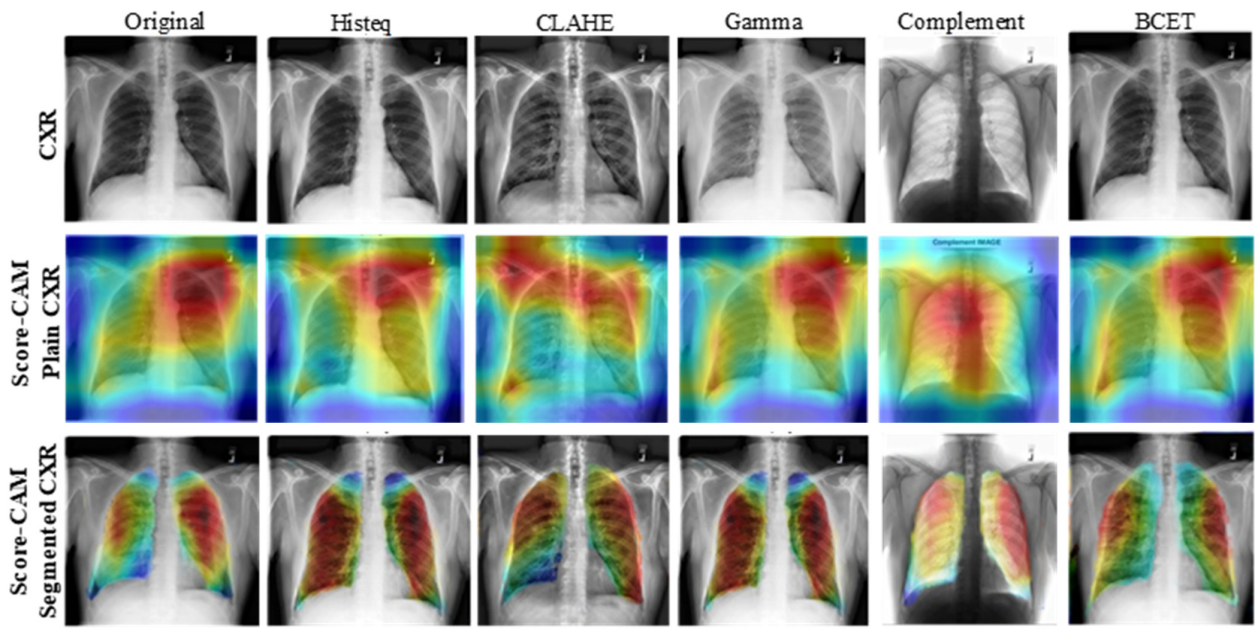

Figure 10. Score-CAM visualization of correctly classified COVID-19 X-ray images using the different enhancement techniques: CXR (top row), Score-CAM heat map on original CXR (middle row) and on segmented lungs CXR (bottom row).

It is clearly visible from the heat map visualization of both the original and segmented lungs using Score-CAM technique is that the decision making in original CXR is not always from the lungs areas (Figure 10). The areas which are mostly contributing to take decision by CNN models are not a part of the lungs always or in most of the cases. It is noticeable that there was no increase in performance with the segmented lungs in comparison to the plain X-rays however it can be seen, that the segmented lungs helped the CNN to take decision from the main region of interest compared to the plain X-rays, which is also criticized in recent papers [28, 76].

It is also interesting to see how the Gamma enhancement technique is outperforming other enhancement techniques for a sample case where almost all the techniques have misclassified a COVID-19 X-rays to either class: Normal or Lung Opacity but Gamma enhancement technique have correctly classified it. It can also be seen from Figure 11 that Gamma enhancement technique on the segmented lungs is taking decisions from the region of interest, i.e. lungs, and correctly classifies the images otherwise which was miss-classified by the network.

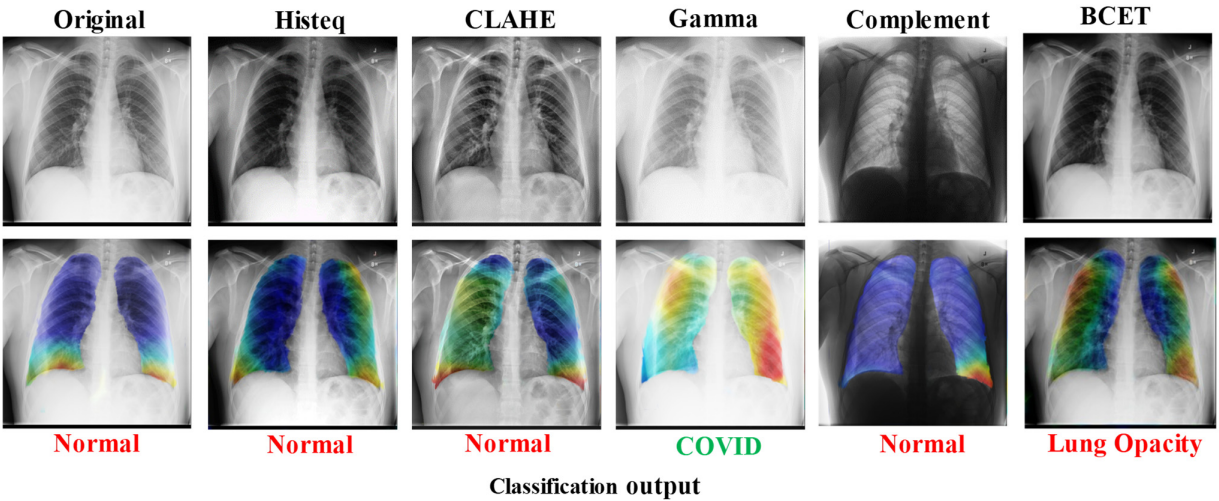

Figure 11. Gamma Enhancement on segmented lungs correctly classify COVID-19 x-ray while others miss classify the sample X-ray images.

## 5. Conclusion

This work explores the effect of different image enhancement techniques in the automatic detection of COVID-19 from the CXR images using deep Convolutional Neural Networks. The performance of six different CNN models for five different enhancement techniques were evaluated for the classification of COVID-19, Non-COVID lung opacity and normal CXR images. ChexNet model with gamma enhancement technique provided the best performance without image segmentation whereas DenseNet201 with gamma enhancement technique outperforms for the segmented lungs.

The classification accuracy, precision and recall for the detection of COVID-19 were found to be 96.29%, 96.28%, and 96.28% without segmentation and 95.11%, 94.55% and 94.56% with segmentation respectively. Score-CAM visualization technique confirms the reliability of the trained models as the decision was made from the lung regions in the segmented CXR images. Thus the results reaffirms the importance of accurate segmentation of lungs from the CXR images, which can assist machine learning models in diagnostic decisions. This deep AI based system can be helpful as a fast screening tool, especially during the pandemic period, and can save the casualties due to delay or not accurate diagnosis.

## Author Contribution

## Funding

Qatar University COVID19 Emergency Response Grant (QUERG-CENG-2020-1) provided the support for the work and the claims made herein are solely the responsibility of the authors.

## Acknowledgments

The publication of this article was funded by the Qatar National Library.